\documentclass[%
 preprint,
 amsmath,amssymb,
 aps,
]{revtex4-2}

\usepackage{graphicx}% Include figure files
\usepackage{dcolumn}% Align table columns on decimal point
\usepackage{bm}% bold math
\usepackage{placeins}
\usepackage{comment}
\usepackage{amsmath}
\usepackage{lipsum}
\usepackage{tikz}
\usepackage{gensymb}
\usepackage{xcolor}
\usepackage{footmisc} % For referencing same footnote multiple times.
\usepackage{relsize} % For changing size of math symbols
\usepackage{nicefrac} % For good looking fractions
\usepackage{dblfloatfix} % fix for bottom-placement of figure
\usepackage{caption} 
\usepackage{tabulary}
\newcolumntype{K}[1]{>{\centering\arraybackslash}p{#1}}

\begin{document}

%\preprint{APS/123-QED}

\title{Taming geometric frustration by leveraging structural elasticity}% Force line breaks with \\

\author{Janav P. Udani}
\author{Andres F. Arrieta}%
 \email{aarrieta@purdue.edu}
\affiliation{%
Purdue University
}%


\date{\today}% It is always \today, today,
\begin{abstract}
Geometric frustration appears in a broad range of systems, generally emerging as disordered ground configurations, thereby impeding understanding of the phenomenon's underlying mechanics. We report on a continuum system featuring locally bistable units that allows for the controlled and self-sustained manifestation of macroscopic geometric frustration. The patterning of the units encodes a finite set of ordered ground configurations (spin-ice states) and a unique family of co-existing higher-order frustrated states (spin-liquid states), which are both activated upon unit inversion. We present a strategy for accessing any globally frustrated state on-demand by controlling the constitutive units’ inversion sequence. This control strategy allows for observing the unfolding of geometric frustration as the microstructural features evolve due to the energy minimization of the constitutive units' interactions. More broadly, our model system offers a blueprint for ``taming” macroscopic geometric frustration, enabling novel applications such as path-driven computation and solving optimisation problems using structural systems. 
\end{abstract}
%%
%%
\begin{comment}

%%
%%

\begin{description}
\item[Usage]
Secondary publications and information retrieval purposes.
\item[Structure]
You may use the \texttt{description} environment to structure your abstract;
use the optional argument of the \verb+\item+ command to give the category of each item. 
\end{description}

%%
%%

\end{comment}
%\keywords{abc; pqr}%Use showkeys class option if keyword
\maketitle

\begin{figure}[t]
%\centering
\includegraphics[width=0.7\textwidth]{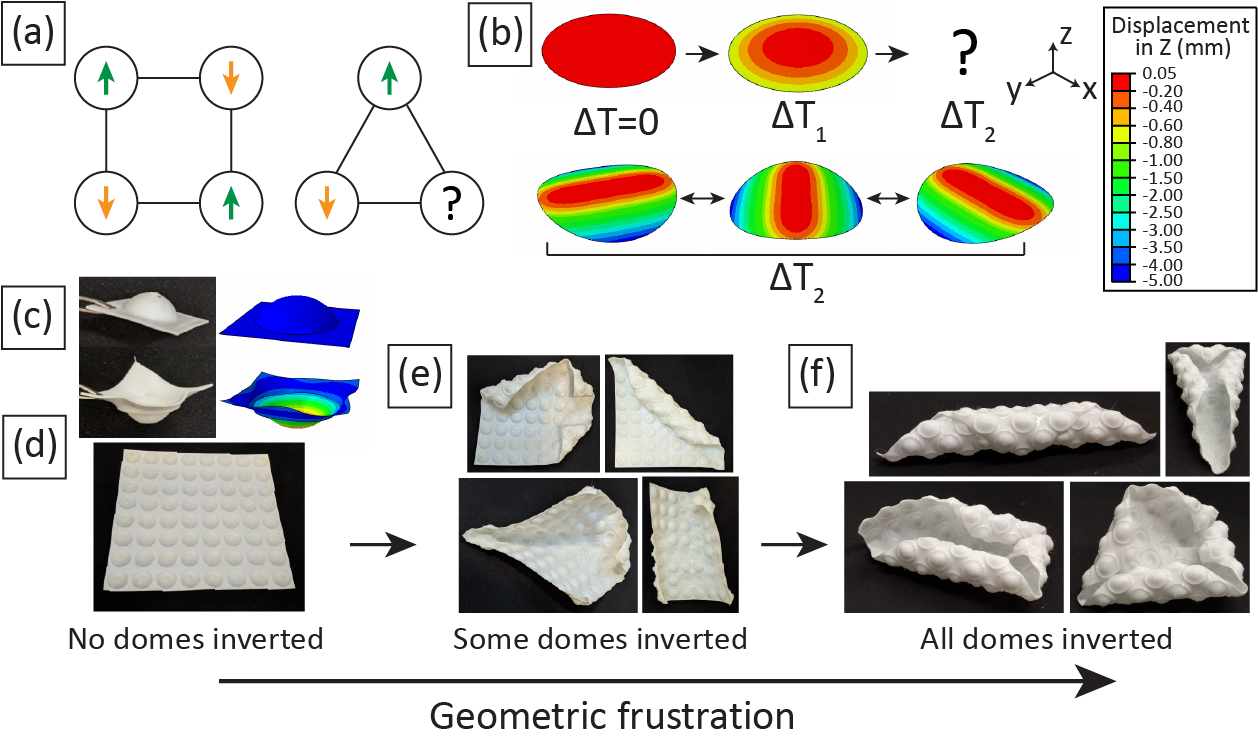}
\caption{ (a) Schematic illustrating the Ising model. Antiferromagnetic interaction spins are satisfied on a square but cannot be satisfied in a triangular arrangement. (b) Structural degeneracy in heated thin plates. When the temperature gradient crosses a threshold value, the plate bifurcates from a spherical to a cylindrical bending configuration with a degenerate bending axis. (c) A dome unit in the base and inverted states (see SI video 1). (d) Square patterned dome metasheet (e) - (f) Manifestation of geometric frustration as an increasing number of domes are inverted.}
\label{fig1}
\end{figure}

\begin{figure*}[t]
\centering
\includegraphics[width=1\textwidth]{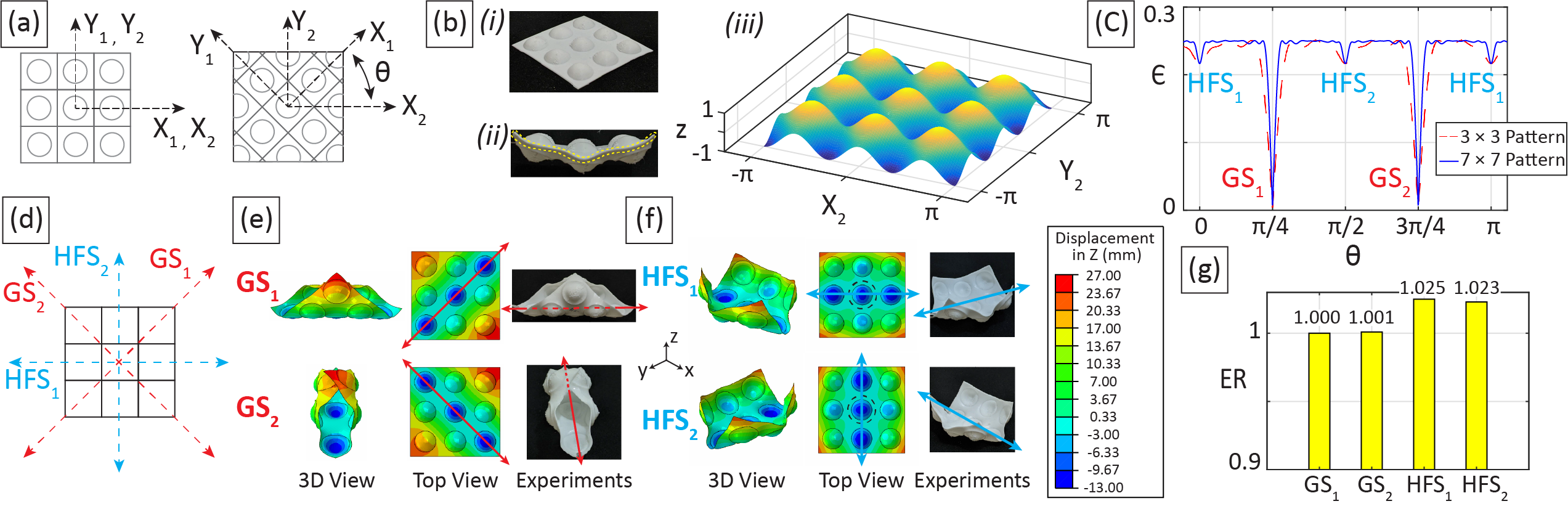}
\caption{(a) Schematic defining the co-ordinate frames employed in the homogenisation model. (b) \textit{(i)} Base state of the 3 $\times$ 3 metasheet. \textit{(ii)} Undulating profile when the domes are inverted \textit{(iii)} Biharmonic function modelling the undulating profile in the inverted metasheet. (c)-(d) $\epsilon(\theta)$ plot and schematic of the metasheet illustrating the minima in the bending stiffness orientations corresponding to the ground states (GS$_{1}$, GS$_{2}$) and higher-order frustrated states  (HFS$_{1}$, HFS$_{2}$). (e)-(f) FE and experimental results illustrating the ground states and the higher-order frustrated states. The top view columns illustrate the deformation field corresponding to the respective state, over-plotted on the base state of the metasheet. The arrows indicate the bending axes for the respective states. (g) Strain energy levels associated with the four hierarchical stable states. The values are normalised as the Energy Ratio (ER) with respect to the strain energy associated with GS$_{1}$.}
\label{fig2}
\end{figure*}

%\tableofcontents
%%
%%
%%
Geometric frustration arises when a lattice system cannot simultaneously minimise all of its local interaction energies due to constraints~\cite{Moessner2006,Han2008}. This leads to a high degree of degeneracy in the system, and the emergence of multiple disordered, high entropy ground configurations. This phenomenon is most commonly seen in ordinary water ice~\cite{Moessner2006}, wherein the hydrogen ions surrounding the central oxygen atom are arranged in a tetrahedral configuration to minimise the interaction energy. Interestingly, geometric frustration is also argued to have implications on the folding of proteins to form a well-defined structure with biological functionality~\cite{Grason2016}. The essence of this phenomenon is captured using a model of Ising spins with antiferromagnetic interactions arranged in a two-dimensional triangular lattice~\cite{Wannier1950,Houtappel1950}. The three interaction energies cannot all be simultaneously minimised, resulting in two antiparallel spins, while the third is frustrated (Fig.~\ref{fig1}a). This simple model is extrapolated to describe several physical systems exhibiting frustration including artificial spin-ice systems~\cite{Ramirez1994,Siddharthan1999,DemHertog2000,Wang2006,Qi2008,Morgan2011,Daunheimer2011}, colloid systems~\cite{Han2008,Leoni2017}, periodically arranged magnetic rotors~\cite{Mellado2012}, acoustic channel lattices~\cite{Wang2017} and elastic structural systems~\cite{Harnett2013,Lechenault2014,Hoon2014}. However, the majority of these systems show high entropy, disordered states, which impede deeper understanding of the mechanics of geometric frustration. Therefore, there is a growing interest in establishing physical systems featuring ordered frustrated configurations, as a route to unveiling the intricacies of important processes such as natural protein folding~\cite{Grason2016}, self-assembly of fibres~\cite{Hall2016,Lenz2017} and nanotubules for targeted drug delivery~\cite{Ramirez2003}, and memory storage in microelectronic devices~\cite{Ramirez2003,Merrigan2020}.\\
\indent Mechanical systems exhibiting deformation-driven frustration present an intriguing avenue for understanding the emergence of order in otherwise disordered systems. Cellular structures featuring beam~\cite{Hoon2014} and shell units~\cite{Seffen2006,Harnett2013,doi:10.1002/advs.202001955} have been shown to naturally exhibit unique and ordered global patterns that emerge either as a result of local elastic constraints or non-near neighbour (NNN) interactions. The mechanics for the emergence of degenerate states in elastic systems were first investigated by Mansfield, in his seminal work employing a prototypical thin plate subjected to a through-thickness temperature gradient~\cite{Mansfield1962,Mansfield1965}. For low thermal field values, the deformed configuration exhibits constant spherical curvature (positive Gauss curvature, $\mathcal K >0$), characterised by both membrane and bending stresses. This spherical deformation results from the absence of any preferential direction for bending in the ideal circular plate. However, upon reaching a threshold in the external field value, the membrane stresses are released. This results in the plate bifurcating into a developable cylindrical form ($\mathcal K=0$), curling up about any diametrical axis (for an ideal plate). This leads to an infinitely many post-buckled degenerate states (Fig.~\ref{fig1}b). Invariably, imperfections in practical realisations control the orientation of the bending axis. Hence, while the possibility of achieving bending in any direction is fascinating and desirable for shape morphing, uncontrolled degeneracy renders it as a prohibitively complicated controls problem. Furthermore, this concept and other examples of magnetic, colloidal and elastic frustrated systems rely on continuously applied external forcing fields for frustration to emerge, thus limiting the utilisation of geometric frustration in practical applications.\\
\indent We report on a lattice system that features controlled macroscale geometric frustration that is primarily driven by the bistability of the microscale units. The system is composed of bistable dome-like units arranged in a continuum metasheet. Local inversion (local buckling) of the domes introduces pre-stress in the system. The interactions between the inverted units’ deformation fields result in global geometric frustration in the lattice (Figs.~\ref{fig1}c-d). We note that the dome inversion and the ensuing geometric frustration are elastic and reversible in nature. Consequently, the emergence of geometric frustration is independent of the material properties as long as the inversion-induced stresses are within the failure limits (see SI section 1, Materials and Methods). The macroscale frustration in our system uniquely manifests in the form of hierarchical multistability~\cite{Seffen2007,doi:10.1002/advs.202001955}, which we define as the emergence of multiple global states for a given dome inversion pattern. Hierarchical multistability departs from the characteristic one-to-one correspondence between the microscale and the macroscale states commonly found in mechanical metamaterials (Figs.~\ref{fig1}e-f, see SI section 2 for length scale definitions). The studied model lattice features four unique characteristics with regards to geometrically frustrated systems $–$ (i) local bistability allows for self-supported pre-stress, thus eliminating the need for sustained external forcing fields to realise geometric frustration; (ii) the elastic constraints imposed by the units patterning naturally lead to the emergence of a finite set of ordered global ground configurations; (iii) in addition to the ordered ground states, the system displays selectively ordered-disordered configurations that closely resemble the spin-liquid states observed in condensed matter frustrated systems; and (iv) all the global hierarchical stable states can be achieved on-demand by controlling the interactions between the units through the history of dome inversions (i.e., the spatiotemporal deformation path). The designs, analyses and results presented in this letter serve as general blueprints for studying geometric frustration at an accessible length scale, as well as enabling unique applications in mechanical computation, soft robotics~\cite{doi:10.1002/advs.202001955} and morphing structures~\cite{Udani2021}.\\
\indent The inversion of local units results in the global out-of-plane bending of the metasheet, thus allowing for reversible 2D to 3D shape and property transformation (Figs.~\ref{fig1},~\ref{fig2}). Similar to Mansfield’s thin plates, the degeneracy in our system arises due to a non-unique cylindrical bending axis once a sufficient number of domes with interacting deformation fields are inverted. Contrary to the heated flat plates' case, the resistance to out-of-plane deformation is not uniform throughout the dome sheet. Instead, the resistance to bending is affected by the metasheet's topology, which is dictated by the patterning of the units. This purely geometric feature leads to a finite set of preferential global bending directions that limit the degree of degeneracy, thereby allowing for the controlled manifestation of geometrically frustrated macroscopic states. To better understand this characteristic, we adopt a homogenisation scheme following Ref.~\cite{Seffen2007}, designed to model the emergence of global modes due to the variation in bending stiffness from dome inversions in the metasheet. We employ a simple square pattern for the analysis and the schematic in Fig.~\ref{fig2}a outlines the relevant co-ordinate frames. The square patterned units are aligned with the $X_{1}Y_{1}Z_{1}$ co-ordinate set; the frame is fixed with respect to the units but is free to rotate by the angle $\theta$ with respect to a globally fixed frame $X_{2}Y_{2}Z_{2}$, such that $X_{1} = X_{2}\cos(\theta) + Y_{2}\sin(\theta)$ and $Y_{1} = -X_{2}\sin(\theta) + Y_{2}\cos(\theta)$. Inverting the domes results in the metasheet adopting a curled configuration with an undulating profile that is modelled by a smooth function $z = f_{1}(X_{1},Y_{1}) = f_{2}(X_{2},Y_{2},\theta)$, capturing the middle points' heights from the $X_{2}Y_{2}$-plane and assumed constant thickness $t$. Without loss of generality, we assume that cylindrical bending of the metasheet always occurs about the $X_{2}$-axis; this assumption does not impose any limiting constraints as the orientation of the domes $\theta$ is set to be free with respect to the $X_{2}$-axis.  Consequently, the bending stiffness can be expressed as,
\begin{equation}
B(Y_{2}) = E \int{z^{2}t} dX_{2},
\label{eq:1}
\end{equation}
where, $E$ is the elastic Young’s modulus of the material. Next, $\kappa(Y_{2})$ is defined as the singly varying curvature of the middle plane, and thus, the angle subtended by the bent metasheet is,
\begin{equation}
\phi = \int{\kappa(Y_{2})} dY_{2} = \int{\frac{M}{B(Y_{2})}} dY_{2} \propto \int{\frac{dY_{2}}{\int{z^{2}dX_{2}}}} = \frac{1}{\epsilon},
\label{eq:2}
\end{equation}
where, $M$ is any general bending moment leading to bending about the $X_{2}$-axis, $\epsilon$ is a relative measure of the bending stiffness and linearity is assumed. With this setup, for any given $z$ profile, the directions of least bending stiffness are identified by analysing the variation in $\epsilon$ as a function of the dome orientation $\theta$. We assume an idealised biharmonic profile $z = \cos(nX_{1})\cos(nY_{1}) = \cos n(X_{2}\cos(\theta) + Y_{2}\sin(\theta))\cos n(-X_{2}\sin(\theta) + Y_{2}\cos(\theta))$, with $X_{2},Y_{2} \in [-\pi,\pi]$ and, $n$ defining the dome packing density, that qualitatively captures the undulating profile for an inverted square metasheet  (Fig.~\ref{fig2}b). The results indicate two well-defined minima in the $\epsilon(\theta)$ plot, thereby suggesting the existence of two unique bending directions when a sufficient number of interacting domes are inverted. Interestingly, both the minimum bending stiffness orientations are aligned with the diagonals, i.e., with the two least packing directions of the square patterned units (Fig.~\ref{fig2}d), this is confirmed by the FE and experimental results (Fig. 2e) (see SI, Materials and Methods for details). This indicates the emergence of $2^{nd}$ neighbour interactions in the frustrated stable states that develop as an increasing number of domes are inverted in the metasheet. 

The observed cylindrical configurations bear close resemblance to the ground state configurations seen in other examples of geometrically frustrated systems – (i) they have bending axes aligned with the programmed preferential directions; (ii) they exhibit identical strain energy levels (Fig.~\ref{fig2}g); and (iii) they are reached when the maximum self-sustaining pre-stress field is introduced. The latter feature implies that inversion of all domes in the metasheet is analogous to freezing down to $0 K$ in naturally frustrated systems. However, in contrast to the random and highly degenerate ground states seen in other systems, the dome metasheets sustain a finite $m-$dimensional manifold of ground configurations, $m$ being the number of least packing directions that are programmed purely by virtue of the unit patterning in the metasheets. Finally, we note the emergence of $2^{nd}$ neighbour interactions in the ordered ground states in our system is not surprising as the mechanism of long range interactions leading to ordered global states has also been found in other classes of geometrically frustrated systems~\cite{Metcalf1974,Daunheimer2011,Hoon2014}.\\
\indent Besides the two global minima in the $\epsilon(\theta)$ plot, there are two additional local minima oriented orthogonally to the unit-patterning directions of the metasheet. To investigate this feature we perturb the ground states in both FE and experiments by applying external mechanical forces and find that the local minima indeed correspond to additional physically stable global states for the same dome inversion pattern, i.e., all domes inverted (Fig.~\ref{fig2}f; see SI video 2). Noticeably, these states exist at a higher strain energy level than the ground configurations (Fig.~\ref{fig2}g). Consequently, these are associated with a shallower potential well in the global energy landscape, and the metasheet is easily perturbed out from these stable configurations. This observation is supported by the fact that the higher energy states feature a shorter bending axis compared to the ground states. Since the bending axis is equal to the side length in the former and equal to the diagonal length in the latter, the metasheet has a higher propensity to assume the more stable ground configurations. Based on the symmetry in the shapes and as observed in the $\epsilon(\theta)$ modelling results, these higher energy states are uniquely characterised by the metasheet being “trapped” in a configuration between the two ground states. The metasheet has equal propensity to go to either of the ground states and in the process, assumes a higher energy frustrated configuration between two (already) frustrated ground states. We term these higher energy states as \textit{higher-order frustrated states}. In a physical sense, these are characterised by the inability of the metasheet to release all of the membrane strain energy, thereby assuming a local minimum displaying stronger stretching-bending interactions. These are energetically unfavourable compared to their ground state counterparts, where the strain energy is primarily bending dominated. Finally, we note that the emergence of hierarchical multistable states is conditional on the inverted units' ability to introduce preferential global bending directions in the metasheet. We find that this feature is directly dependent on the packing density of the domes. We predict the associated design limits necessary for the manifestation of geometric frustration in our metasheet using an analytical approach modelling the topology of an inverted unit, as detailed in SI section 3.\\
\begin{figure*}[ht]
\centering
\includegraphics[width=1\textwidth]{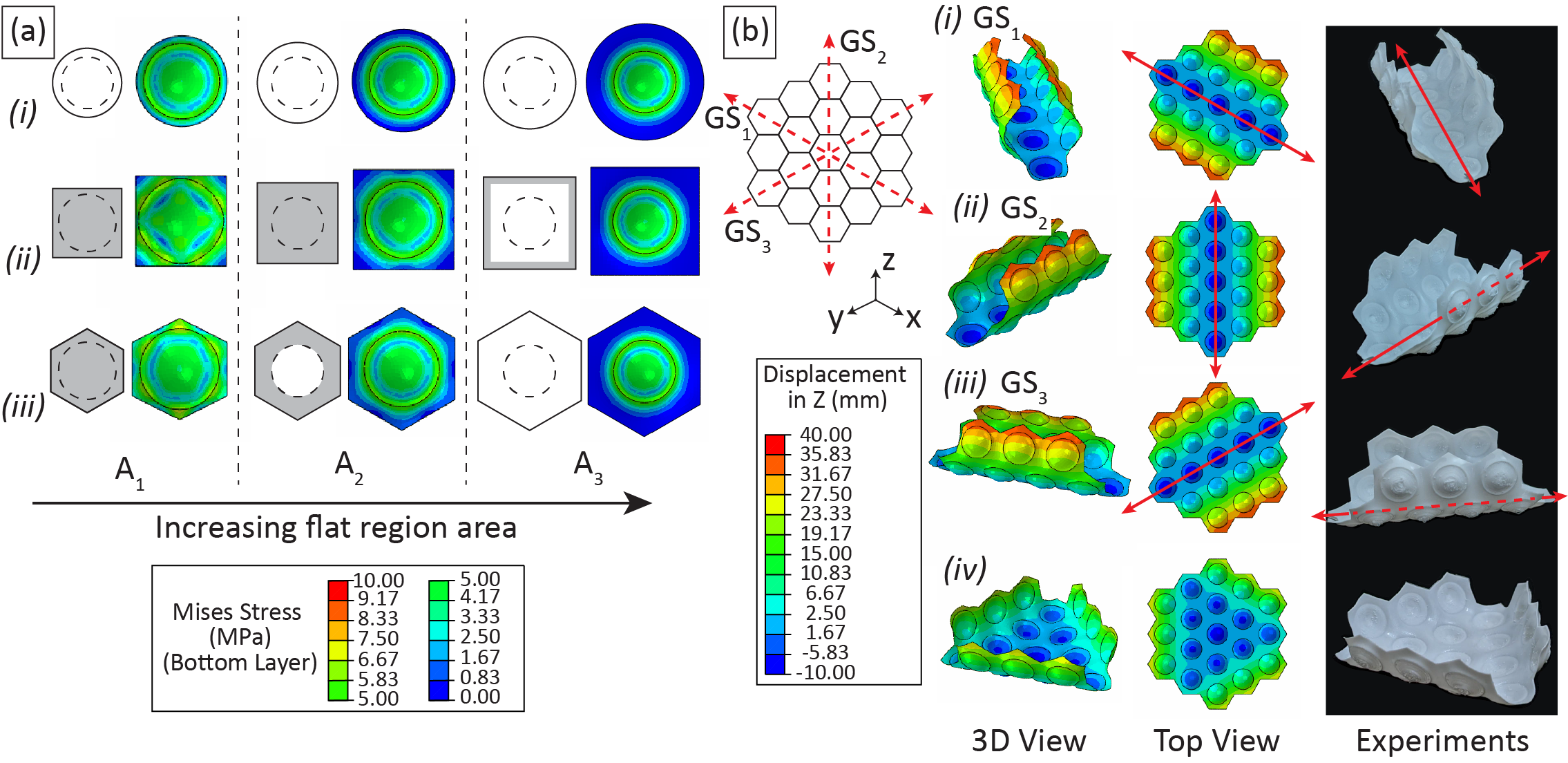}
\caption{(a) Higher-order post-buckling shapes for units featuring a \textit{(i)} circular \textit{(ii)} square and \textit{(iii)} hexagonal boundary shape. The grey shaded area on the schematics indicates the region featuring a non-axisymmetric deformation state. The central dome reverts back to its preferred mirror-buckled state as the flat planform area increases or the number of symmetry axes in the boundary shape are increased. (b) The three bending axes and \textit{(i)}-\textit{(iii)} the associated ground states for a hexagonal metasheet. \textit{(iv)} The higher-order frustrated state for the hexagonal metasheet displays a non-developable positive Gaussian curvature surface when all domes are inverted (see SI video 3). The top view column illustrates the deformation field corresponding to the respective state, over-plotted on the base state of the metasheet. }
\label{fig3}
\end{figure*}
\indent An interesting consequence of the multiple global frustrated states is the emergence of local higher-order buckling modes on the individual domes~\cite{Fitch1968,Taffetani2018}. This feature is distinctly seen on the centre dome in the higher-order frustrated states (black dashed circle in Fig.~\ref{fig2}f), whereby the centre dome departs from the locally axisymmetric (mirror-buckled) configuration to assume an elliptical shape with major axis aligned with the side of the square (see SI section 4). These local post-buckled shapes are again characterised by strong stretching-bending interactions, in contrast to the mirror-buckled state where the deformation is primarily bending dominated~\cite{Vaziri2008}. The emergence of local post-buckled shapes in our system is primarily dictated by two parameters, (i) the individual unit design, a microscale property; and (ii) the global configuration that the system assumes following unit cell inversion, which is a macroscale property. Investigating the microscale design, we find that the flat's region geometry surrounding the dome plays a crucial role in the manifestation of local post-buckled shapes. The units' boundary shape largely dictates the elastic interactions with the neighbouring inverted units (or in general any boundary conditions). When the interactions are strong, they impose a higher-order polygonal post-buckled shape on the central dome, moving away from the preferred axisymmetric inverted shape. This characteristic is visualised in a simple experiment involving units with different geometries but equal flat region area (Fig.~\ref{fig3}a). We pin the boundary edges to simulate the limiting case of the elastic constraints that are imposed on any individual unit in the metasheet when sufficient domes are inverted. The results indicate that for non-axisymmetric boundaries and a small flat region area, the edge constraints impose a non-axisymmetric post-buckled shape on the central dome in the inverted configuration. The break in axisymmetry originates from the boundary shape and is found to localise away from the dome centre as the flat region area increases. Furthermore, the central dome regains an axisymmetric shape for smaller area values if the boundary shape features a higher number of symmetry axes (e.g., for a hexagon as compared to a square, Figs.~\ref{fig3}a.\textit{(ii)}-\textit{(iii)}). In the limiting case of a circular planform, axisymmetry is never lost. This emergence of microscale higher-order buckling modes based on the unit geometry can be subsequently leveraged for programming the desired number of hierarchically multistable ground states at the macroscale. As illustrated in Fig.~\ref{fig2}, a regular square pattern of the units features two preferential global bending directions. Similarly, a regular hexagonal pattern features three least packing directions due to the unit's symmetry, thus showing three degenerate global ground states after a sufficient number of units are inverted (Fig.~\ref{fig3}b). Consequently, the microscale geometry can be tuned to program the desired number of macroscale ground states. However, triangles, squares and hexagons are the only polygonal geometries that can be arranged in a regular, completely packed 2D pattern. All other tessellations form irregular patterns and introduce broken symmetries, but in the process enable the ability to program non-uniform directions of least packing resulting in unique, irregular global shapes.\\
\begin{figure*}[h!]
%\centreing
\includegraphics[width=1\textwidth]{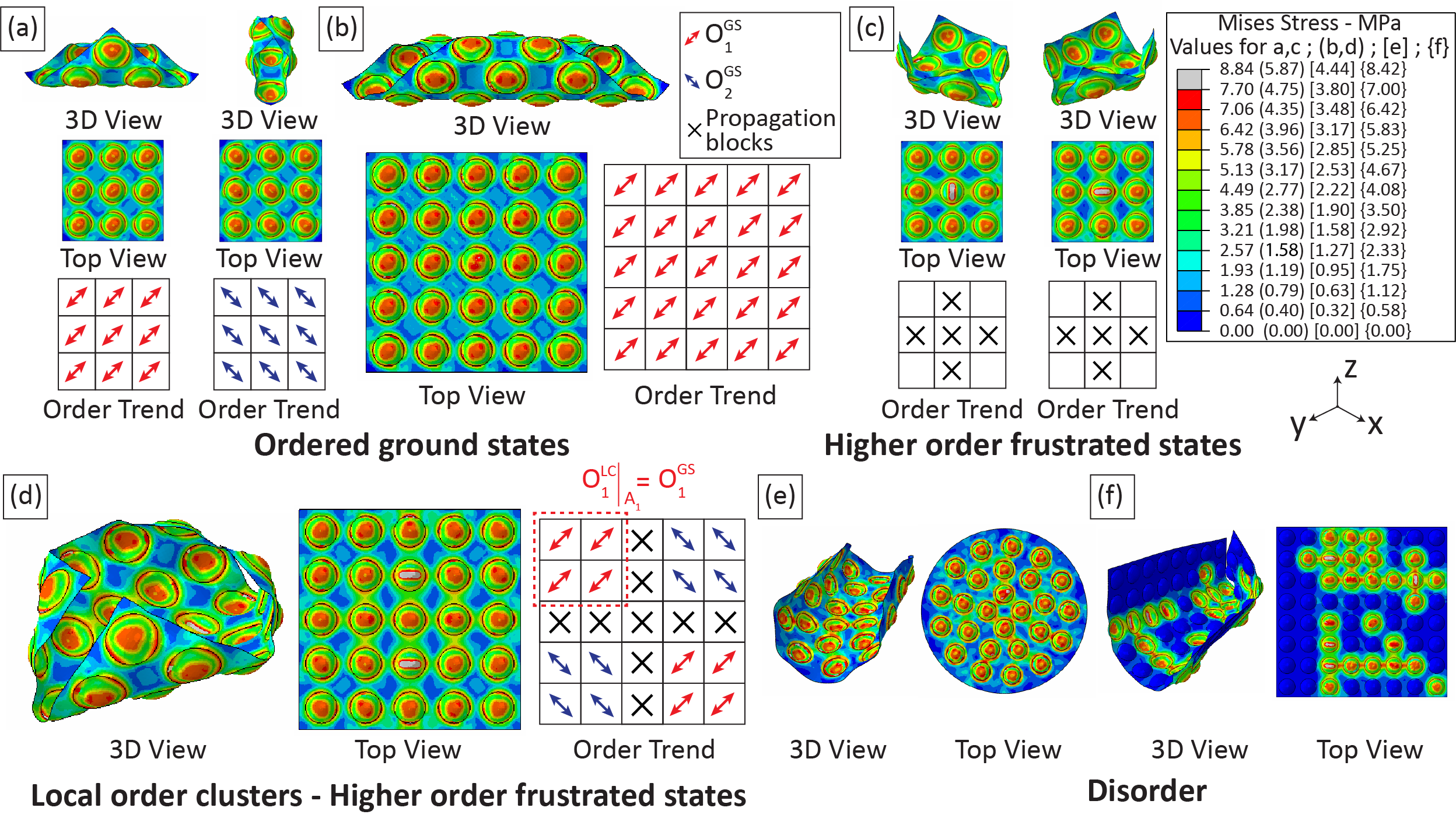}
\caption{(a)-(b) Ordered ground states ($O^{GS}$) featuring ``concerted or aligned local buckling" of the units for a $3 \times 3$ and $5 \times 5$ square metasheet (see SI section 5 for experimental images of the $5\times 5$ metasheet). (c)-(d) Local order clusters ($O_{m}^{LC}$ example highlighted by the dashed red square in (d)) and non-developable deformation blocks preventing order propagation in the higher-order frustrated states $O_{n}^{HFS}$ for the respective metasheets. (e) Disorder within the microstructure achieved by random dome patterning and (f) when an insufficient number of interacting domes are inverted in the metasheets. The top view figures illustrate the deformation field corresponding to the respective state, over-plotted on the base state of the metasheet.}
\label{fig4}
\end{figure*}
\indent Along with the shape of the unit boundary, the metasheet's global deformed state also determines the post-buckling behaviour of the individual domes. In general, the global ground states are characterised mainly by “concerted local buckling” of the microscale units. This is illustrated in Figs.~\ref{fig4}a-b for square patterned metasheets. In these cases, the units assume elliptical post-buckled shapes with major axes diagonally aligned. Interestingly, this alignment direction coincides with the bending axis associated with each global ground state. Thus, there is an underlying order $O^{GS}$ for the (local) post-buckling behaviour that propagates across all the units in the global ground states. A square patterned metasheet exhibits two ground state ordering trends, $O^{GS} \in [O_{1}^{GS},O_{2}^{GS}]$, 
corresponding to the order associated with each preferential direction as programmed by the unit arrangement. The units' tendency to naturally and collectively orient towards their NNN in the post-buckled configuration again confirms the role of long-range interactions in the emergence of ordered global states, as previously discussed in the homogenisation model results (Fig.~\ref{fig2}). In contrast, analysing the ordering trends in the units' post-buckling behaviour for the higher-order frustrated states, $O^{HFS}$, no such long-range order similar to $O^{GS}$ is evident. Instead, we find spatially distributed clusters of local order throughout the metasheet (Figs.~\ref{fig4}c-d). This is mathematically expressed as $O^{HFS} \in [O_{1}^{HFS}, O_{2}^{HFS},...]$ , where, $O_{n}^{HFS} = O_{1}^{LC}\Big|_{A_{1}} + O_{2}^{LC}\Big|_{A_{2}} + ... $. Here, each $O_{m}^{LC}$ corresponds to a different order cluster. The ordering trend of the resulting $n^{th}$ higher-order frustrated state $O_{n}^{HFS}$ is determined by combining the covered areas $A_{m}$ of each individual cluster $O_{m}^{LC}$. While the set of local order clusters can in principle be very large, we note that the microstructural post-buckling shapes are still constrained by the unit design and the pattern-driven finite number of preferential bending directions encoded into the metasheet. Further inspection of the results in Figs.~\ref{fig4}c-d reveals that order trends resembling both $O_{1}^{GS}$ and $O_{2}^{GS}$ are evident, albeit at different spatial locations for distinct groups of units. Neither of the ground state ordering trends has enough authority to force the metasheet into its respective global configuration. 
This results in deformation along diagonally opposite corners in a manner resembling both the ground states. However, the units in the centre of the metasheet are trapped with equal propensity to adopt either state. As a result, the metasheet's central region assumes a locally non-developable deformation state impeding order propagation throughout. This non-developable region characterises all the higher-order frustrated configurations and is essential for their emergence. \\
\indent Interestingly, the appearance of local order clusters is qualitatively reminiscent of the spin-liquid regime found in natural, temperature-driven frustrated systems~\cite{Moessner2006,Balents2010}. When such systems are cooled below the Curie-Weiss temperature, the spins form into local clusters with strong interactions that obey the ground state constraints, albeit, there is no long range propagation of the associated order within the system. This is the precise characteristic that we find in the higher-order frustrated states in our dome metasheets. However, while the spin-liquid regime appears at a distinct temperature range different from the spin-ice regime (ground-state configurations) in condensed matter frustrated systems, we uniquely find that stable configurations qualitatively resembling both of these regimes can co-exist for the same input conditions on our system. For example, when all the domes are inverted, we find both the developable ground states (ordered spin-ice states) and the higher-order frustrated states (spin-liquid states) as illustrated in Figs~\ref{fig4}a-d. With this characteristic, our system joins an exclusive group of exotic systems exhibiting the co-existence of these two distinct stability regimes~\cite{Schiffer1994,Petrenko1998,Mirebeau2002,Sen2015}. The appearance of both families of states in our system is a purely elastic consequence (compatible with different material chemistries, see SI section 6) determined by the programmed preferential bending directions and how much of the stored membrane energy -- added by virtue of dome inversion at the microscale -- is released at the macroscale. In essence, this co-existence of two qualitatively distinct stability regimes for the same input conditions is a result of the intriguing interactions at play between structural elasticity and geometric frustration in our continuum lattice system.\\
\indent Finally, local disorder can also be programmed into the microstructure by randomly patterning the domes in the metasheet (Fig.~\ref{fig4}e). In such cases, inverting all domes results in global shapes with no clearly defined preferential bending directions. These degenerate global shapes feature a microstructure consisting of inverted domes that assume random and strongly disordered post-buckled shapes. In these designs, the interactions between structural elasticity and geometric frustration are not meaningfully leveraged, yielding macroscale shapes that while unique, are neither robust nor easily controllable. Additionally, disordered configurations in the microstructure also emerge when an insufficient number of interacting domes are inverted, resulting in limited authority to impose the programmed shape on the macrostructure (Fig.~\ref{fig4}f). However, in general, we find that by leveraging the interactions between elastic constraints and deformation-driven geometric frustration, the prevalent behaviour is ordered configurations, whereas disordered configurations need to be intentionally programmed into the system.\\
\begin{figure*}[t]
%\centering
\includegraphics[width=1\textwidth]{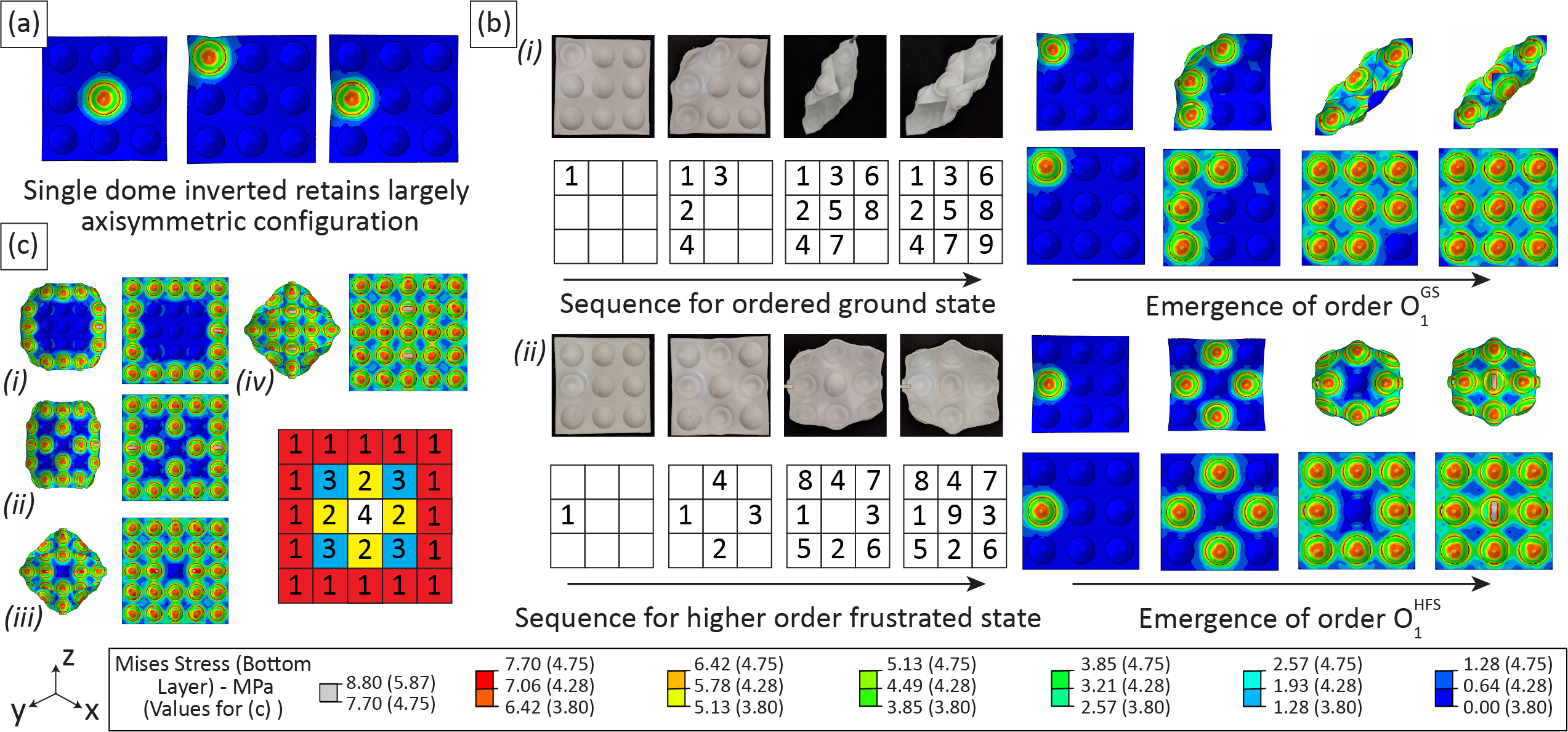}
\caption{(a) Single inverted dome retains a largely axisymmetric configuration. (b) Different dome inversion sequences leading to the desired global frustrated states for the $3 \times 3$ square metasheet. \textit{(i)} Diagonal inversion sequence leading to the ordered ground state. \textit{(ii)} Diamond-like symmetric inversion sequence leading to the higher-order frustrated state. (c) Dome inversion sequence \textit{(i)} - \textit{(iv)} leading to a higher-order frustrated state for the $5 \times 5$ square metasheet. All the FE simulation results are displayed in the top view. For (b) and (c), the plot pairs for each state correspond to the top view of the deformed configuration and the top view of the deformed field over-plotted on the base configuration.}
\label{fig5}
\end{figure*}
\indent As noted before, the post-buckled shape and orientation of the units characterises the global frustrated states. These characteristics can be further leveraged to achieve on-demand access to any desired global frustrated state. A dome unit is stress-free in its base state and assumes a stressed configuration only upon inversion. When a single unit is inverted, it assumes the axisymmetric mirror-buckled shape (Fig.~\ref{fig5}a). However, as an increasing number of interacting units are inverted, the domes progressively transition into one of their polygonal post-buckled shapes depending on the sequence in which the interactions and elastic constraints are imposed. This feature can be leveraged to achieve any desired order, i.e., $O_n^{GS}$ or $O_n^{HFS}$, within the microstructure. Collectively, these interactions force the macrostructure to assume the specific global frustrated state corresponding to the respective microscale ordering scheme. Thus, distinct global frustrated configurations corresponding to the same inversion pattern can be readily accessed simply by controlling the dome inversion sequence. This concept is illustrated on a prototypical $3 \times 3$ square patterned metasheet. When the dome inversion sequence is initiated in a diagonal sense, i.e., inverting a series of $2^{nd}$ neighbour domes  from one corner to the other, the metasheet assumes the corresponding ground state configuration (Fig.~\ref{fig5}b.\textit{(i)}, see SI video 4). In constrast, if the dome inversion sequence is initiated symmetrically about the centre dome, implying we first invert the four middle domes in a diamond fashion and subsequently invert the four corner domes, we find that the diagonally-aligned bending axis necessary for reaching either of the ground states is not activated as the central dome is still in its base, stress-free state (Fig.~\ref{fig5}b.\textit{(ii)}). This inversion sequence imposes a configuration symmetrically aligned with either of the ground states, thus leading to the higher-order frustrated state when the centre dome is finally inverted (see SI video 5). Similar inversion sequences can also be realised for larger metasheets by selectively imposing the desired microscale interactions between units; an example inversion sequence leading to a higher-order frustrated state for a $5 \times 5$ square metasheet is shown in Fig.~\ref{fig5}c (also see SI video 6). Noticeably, this strategy reveals that the mapping of the number of inversion sequences to the total number of frustrated states is not unique; the number of sequences is $N!$, $N$ being the number of domes, and is presumably larger than the set of global frustrated states. While there may be multiple paths leading to a given global state, we note that no single path can guide the system into two frustrated states. This observation renders the inversion sequencing strategy as both robust and reliable in controlling the manifestation of geometric frustration in our system. In essence, we leverage the elastic constraints imposed by the domes' local bistability as a mechanism for selectively reaching the desired microscale interactions needed to realise distinct global frustrated states. This is a powerful concept that enables our system to serve as a test-bed for accessibly studying the mechanisms governing geometric frustration, as well as for opening avenues for novel applications like path-driven computation paradigms in structural systems~\cite{doi:10.1002/advs.202001955}.\\
\indent More generally, the ability to ``tame" geometric frustration enables the design of machine-like structural systems that serve as physical embodiments of complex optimisation problems. The co-existing hierarchically multistable states discretize the system's configuration space into distinct, albeit finite minimum energy points. In this regard, the ground (ordered spin-ice) and higher-order frustrated (spin-liquid) states correspond to global and local minima, respectively. The ability to separate and achieve any global frustrated state on-demand as controlled by distinct sequences of discrete external events (i.e., the local dome inversions and their history) sets a physical paradigm for mechanical optimisation and information processing. In this regard, the presented dome metasheets open an avenue for implementing discrete optimisation paradigms, both exact algorithms like the branch and bound~\cite{Lawler1966Branch-And-BoundSurvey} and metaheuristics like simulated annealing~\cite{Kirkpatrick2007OptimizationAnnealing}, into mechanical metamaterials. The macroscale deformation enables in situ monitoring of the optimisation process or unfolding of geometric frustration, whereby iterative dome inversions allow for reaching the global minimum or ground state if required. The ability to encode the desired number of ground states (global minima) further enriches the optimisation and computation paradigm afforded by our metamaterial architecture and enables the physical mapping of Pareto-optimality. These features are absent in the general class of frustrated systems that exhibit uncontrolled degeneracy, and consequently, inhibit the ability to achieve a desired frustrated state on demand. Contrary to unfrustrated systems with a single global energy minimum, hierarchical multistability enables the mapping of richer optimisation paradigms with complex energy landscapes.\\
\indent Notably, our system naturally exhibits non-volatile memory at the microscale, the local bistable states of the units can be ascribed bit values ``0" and ``1"~\cite{Treml2018b,Chen2021AMemory}. This feature, complimented with the intrinsic deformation-driven optimisation paradigm in our metasheets provides a unique information processing capability on mechanical metamaterials. These characteristics paired together enable our geometrically frustrated metasheets to display a finite-state machine type behaviour that solves optimisation problems based on external inputs, internal unit states (bit values), and also stores long-term event history memory. We envision our system as a blueprint for achieving controlled interactions between structural elasticity and geometric frustration, leading to novel metamaterials with path-driven morphological computation capable of parsing spatially complex mechanical information for soft robotics~\cite{Kofod2006,Rus,Raney2016,Ferrand2019}, haptic devices~\cite{Homberg2015HapticGripper,Li2019Bio-inspiredInterfaces}, wearables, and adaptive aerospace systems~\cite{OlympioK.R.2010,Runkel2019}. 
\section*{Acknowledgements}
The authors acknowledge the support of the NSF-CAREER award No. 1944597 and the DSO DARPA-NAC award, contract No. HR00112090010.

\bibliography{references}

\end{document}